\documentclass[5p,authoryear,hidelinks]{elsarticle}  

\usepackage{subfig}
\usepackage{amsmath}
\usepackage{amssymb}
\usepackage{amsbsy}  
\usepackage{graphicx}
\usepackage{url}
\usepackage{color}
\usepackage[breaklinks]{hyperref}

\newcommand{\nitrate}{ {NO}$_3^{-}$ } 

\def \Eq#1{Eq.\,(\ref{#1})}
\def \Fig#1{Fig.\,\ref{#1}}
\def \Tab#1{Tab.\,\ref{#1}}

\def \f#1#2{\frac{\textstyle #1}{\textstyle #2}}
\def \p#1#2{\frac{\partial #1}{\partial #2}}

\def \vc#1{\boldsymbol{#1}}  

\begin{document}

\date{August 26, 2020}

\begin{frontmatter}

\title{A modelling study of hydrodynamical and biogeochemical processes within the  
California Upwelling System}


\author[1]{Karsten Alexander Lettmann\corref{cor1}}
\ead{k.lettmann@uni-oldenburg.de}
\author[1]{Florian Hahner}
\author[1]{Vanessa Schakau}
\author[1,2]{Tim W\"ullner}
\author[1]{Cora Kohlmeier}

\cortext[cor1]{Corresponding author}

\address[1]{Institute for Chemistry and Biology of the Marine Environment,
Carl von Ossietzky University Oldenburg, Germany}
\address[2]{OFFIS -  Institute for Information Technology, Oldenburg, Germany}



\begin{abstract}
The ROMS modeling system was applied to the California Upwelling System (CalUS) to 
understand the key hydrodynamic conditions and dynamics of the nitrogen-based ecosystem 
using the NPZD model proposed by \citet{Powell_2006}. A new type of sponge layer 
has been successfully implemented in the ROMS modelling system in order to stabilize the 
hydrodynamic part of the modeling system when using so-called "reduced" boundary 
conditions.
The hydrodynamic performance of the model was examined using a tidal analysis based on 
tidal measurement data, a comparison of the modeled sea surface temperature (SST) with 
buoy and satellite data, and vertical sections of the currents along the coast and the 
water temperature. This validation process shows that the hydrodynamic module used in this 
study can reproduce the basic hydrodynamic and circulation characteristics within the 
CalUS.
The results of the ecosystem model show the characteristic features of upwelling regions 
as well as the well-known spotty horizontal structures of the zooplankton community. The 
model thus provides a solid basis for the hydrodynamic and ecological characteristics of 
the CalUS and enables the ecological model to be expanded into a complex ecological model 
for investigating the effects of climate change on the ecological balance in the area 
investigated.

\end{abstract}



\end{frontmatter}



\section{Introduction}

Eastern boundary upwelling (EBU) system  belong to the most productive 
regions of the word ocean, which is due to the fuelling of the photic zone by cool and 
nutrient rich water masses from below based on offshore Ekman transport in surface waters.  
These upwelling regions account for only about 1 \% of the global ocean, but produce about 
20 \% of the global fish catch and are also known to support sea birds and mammals such as 
whales and seals (see e.g. \citet{Kaempf_Chapman:2016} for a general overview of 
global upwelling systems). The four main eastern boundary systems are those off a) California / Oregon / 
Washington in the North Pacific, b) Peru and Chile in the South Pacific, c) off northwest Africa and Portugal 
in the North Atlantic, and d) off South Africa and Namibia in the South 
Atlantic. Apart from these four major systems, a number of other upwelling systems exist 
throughout the global ocean, some of which are year-round features, whereas others occur 
on a seasonal basis \citep{Kaempf_Chapman:2016}.

Understanding the physical and biogeochemical processes in these upwelling systems is of 
great importance. And coupled modelling systems have been valuable tools in the past to 
contribute to this understanding. Within this manuscript, we want to focus on the 
California Upwelling System  (CalUS) and have developed a coupled modelling system for 
that region. In detail, we use a 3D coastal ocean 
circulation model coupled to a lower trophic level nitrogen-based ecosystem model, which 
are part of the ROMS modelling system \citep[\textbf{R}egional 
\textbf{O}cean \textbf{M}odelling \textbf{S}ystem, see e.g.][]{Haidvogel:2000, 
Wilkin:2005}.  

The ROMS modelling system has been applied to the CalUS many times before \citep[see 
e.g.][]{Gruber_al:2006, Song_al:2011, Jacox_al:2014}, in order to study different physical 
and biogeochemical processes. For example, the strong horizontal nutrient gradients and 
lateral horizontal transports by filaments and mesoscale eddies, which are characteristic 
for EBUs, was nicely illustrated \citep[see e.g.][]{Marchesiello_2003, Nagai_2015}.


This manuscript describes the application of the ROMS modelling 
system to the CalUS. The used lower trophic level nitrogen-based ecosystem model is 
based on the  four-component NPZD model by \citet{Powell_2006}, which itself is mainly 
based on the studies by \citet{Spitz_2003} and \citet{Newberger_2003}. We show some 
validation of the physical and biological module. 

The interested reader will find a short overview of the California Current 
System in Section \ref{sec:california_current_system}. In Section 
\ref{sec:model_description}, the modelling System is described, and its validation is 
presented in Section \ref{sec:modelling_system_validation}. Finally, the new sponge layer 
type used to stabilize the ROMS modelling system when using so-called reduced boundary conditions is presented 
and discussed in \ref{sec:sponge_layer}.



\section{The California Current System}
\label{sec:california_current_system}

As the  hydrography and its variability of the California Current System (CCS) has been described in the 
past by many authors 
\citep[see e.g.][and references therein]{HICKEY:1979, Lynn_Simpson:1987, STRUB_JAMES:2000, 
Centurioni_al:2008, CHECKLEY_BARTH:2009, GANGOPADHYAY_al:2011, Kaempf_Chapman:2016}, we only want to give a 
brief description of the CCS in order to provide the background for evaluating the hydrodynamic model features 
presented below.

The California Current System consists of different current features with different water mass 
characteristics due to their source regions, that are located at surface or below surface, and 
which might show a northward or southward net flow structure (see e.g. \citet{CHECKLEY_BARTH:2009} Fig. 1 or
\citet{GANGOPADHYAY_al:2011}  Fig. 3 for a general overview of the different current features).  It extends, 
in the north, from the Transition Zone (50$^\circ$N, separating the North Pacific and Alaska gyres), where 
the east-flowing North Pacific Current \citep[also called the West Wind Drift, see e.g.][]{STRUB_JAMES:2000} 
approaches North America,  to subtropical waters off Baja California, Mexico ($\sim$ 15-25$^\circ$N) in the 
south \citep{HICKEY:1979, CHECKLEY_BARTH:2009}. \citet{CHECKLEY_BARTH:2009} and \citet{GANGOPADHYAY_al:2011} 
summarize diverse current features and processes on different spatial and temporal scales that occur in the 
CCS: wind-driven upwelling, the geostrophically balanced California Current (CC), the coastal jet, the 
California Undercurrent (CU), Inshore Countercurrent (ICC) (Davidson Current), jets (narrow high-speed flows) 
in general, squirts (localized energetic off-shelf flows), filaments, mushroom-head vortices, mesoscale and 
sub-mesoscale eddies, and finally large meanders. When describing the horizontal position of these 
hydrodynamic features, we follow \citet{Lynn_Simpson:1987}, who separate the CCS into an offshore oceanic 
zone ($\approx$ 300 - 1000 km), a near-shore coastal zone ($\approx$ 0 - 200 km) and an intervening 
transition zone ($\approx$ 200 - 300 km) \citep[this spatial division is also evident in Fig. 1 
of][]{CHECKLEY_BARTH:2009}. These zones interact with each other by various mechanisms, and their widths are 
only a rough estimate and are not static.

Concerning the  California Current (CC), the classical view depicts a slow and  broad current 
that flows equatorward within about 1000 km of the west coast of North America connecting the eastward North 
Pacific Current  at approximately 50$^\circ$N to the westward North Equatorial Current at approximately 
20$^\circ$N \citep{STRUB_JAMES:2000, CHECKLEY_BARTH:2009}. It is  a year-round, and surface-intensified flow 
usually in the upper 500~m that carries about 10 Sv \citep{Sverdrup_1942, CHECKLEY_BARTH:2009}, and, 
according 
to \citet{Lynn_Simpson:1987}, the main core of the CC is located within the transition 
zone. However, this 
picture of the slow and broad current has changed  over the last decades \citep[see 
e.g.][]{Davis:1985, HUYER:1998, Centurioni_al:2008, Marchesiello_2003}. According to 
\citet{CHECKLEY_BARTH:2009} and references therein, the southward flow can be partly organized in form of 
intense equatorward jets that are embedded within the region of slower southward flow 
\citep{MOOERS_Robinson:1984, HUYER:1998}. The intense jets have widths of 50 - 75 km,  speeds in excess of 
0.5~m~s$^{-1}$, comprise up to half of the total CC transport, and are mainly located in or near the 
transition zone mentioned above. According to \citet{COLLINS:2003} and looking at graphs in 
\citet{STRUB_JAMES:2000}, the observed California Current jets in and near the transition zone can be seen as 
the inshore edge of the broader mean seasonal equatorward flow of the California Current.  In some cases, 
these jets can be traced back to coastal upwelling jets that separate from the coast, merge offshore (to about 
130$^\circ$W) to become a free, open-ocean jet that maintains its identity as the CC core during spring and 
summer \citep{BARTH_al:2000, STRUB_JAMES:2000, GANGOPADHYAY_al:2011}.

The near-shore coastal zone and the transition zone show a very complex dynamic, which changes during the 
year, and with more mesoscale features present in late summer to early fall
\citep[e.g.][]{STRUB_JAMES:2000}. Within these zones, the upwelling regions, the coastal 
jet, the California Undercurrent (CU), as well as the Inshore Countercurrent (ICC) are located 
\citep{CHECKLEY_BARTH:2009}. It is the source region of coastal and westward propagating 
cyclonic and anticyclonic mesoscale eddies \citep{Kurian_al:2011} and it is the region 
with all those smaller-scales, high-energetic features like those filaments mentioned 
above. In addition, within the coastal zone, there exist some frequent standing eddies, like the 
counterclockwise Southern California eddy located south of Point Conception \citep{CHECKLEY_BARTH:2009}, a 
counterclockwise eddy off San Francisco and about half the distance to Point Conception \citep{HICKEY:1979}, 
eddies near Point Arena, and the Cape Mendocino eddy \citep{Hayward_Mantyla:1990}.

The coastal jet is generated in geostrophic balance due to both a drop in coastal sea 
level and the presence of the cold water front near
the coast \citep{CHECKLEY_BARTH:2009}, which generates a strong, equatorward coastal 
upwelling jet with speeds of up to 1~m~s$^{-1}$. According to the thermal-wind relation, the 
coastal upwelling jet is vertically sheared, with strongest currents near the 
surface, because temperature, salinity and, hence, density vary in the cross-shelf 
direction. The jet is also horizontally sheared and fastest near the strongest cross-shelf 
density difference, i.e. the coastal upwelling front \citep{CHECKLEY_BARTH:2009}.  
Interactions of the alongshore flow with coastal and bottom bathymetric features (capes, 
banks, canyons), in combination with hydrodynamic instability, also leads to intense 
alongshore variability e.g. visible in transient and even persistent meanders 
and eddies especially from spring to early 
fall \citep{STRUB_JAMES:2000,  Centurioni_al:2008, CHECKLEY_BARTH:2009, Drake_al:2011}.

Within the coastal zone of the CCS, two narrow poleward flowing boundary currents are found. 
These currents, the Inshore Countercurrent (ICC) and the California Undercurrent (CU), 
are distinguished from each other by their water mass characteristics, their vertical location, and their 
temporal presence during the year \citep{COLLINS_al:2000}. 

The CU appears as a subsurface maximum of flow between 100 and 250 m depth over the continental slope and 
transports warm, saline equatorial waters \citep{Chelton:1984, Lynn_Simpson:1987, HICKEY:1998}. It is 
considered to originate in the eastern equatorial Pacific and to flow poleward along the North American coast 
\citep{Sverdrup_1942, Lynn_Simpson:1987}. Thus, the CU can be seen as an example of poleward undercurrents 
also present in other major ocean basins, which are usually found over the continental slope and which have 
typical alongshore speeds of 0.1 - 0.3 m~s$^{-1}$ and a depth range of 100 - 300 m \citep{PIERCE_al:2000}. In 
the 
mean, the main core of the CU is located  at 250~m depth near the continental slope and it does not extend 
beyond 100 km from the coast, although there is some seasonal variability of the height and strength (mean 
$\approx$ 5 - 10 cm~s$^{-1}$) of the core \citep{Lynn_Simpson:1987,  COLLINS_al:1996}. In detail, the CU has 
been observed at locations ranging from Baja California  to Vancouver Island \citep{HICKEY:1998}, and 
shipboard surveys along the West Coast of the U.S. show poleward flow over the upper slope at all latitudes 
\citep{PIERCE_al:1996, COLLINS_al:2000}.

According to \citet{COLLINS_al:2000}, the ICC has been reported as a seasonal flow, appearing 
in fall and winter \citep{Reid_Schwarzlose:1962, Lynn_Simpson:1987}. 
It is found over both the shelf and slope and transports shallow, upper ocean waters, which mainly are 
derived from CC waters with some modification by coastal processes. North of Point Conception, the ICC is
sometimes called the Davidson Current or the Davidson Inshore Current \citep{Reid_Schwarzlose:1962, 
HICKEY:1979}.


\section{Description of the coupled modelling system}
\label{sec:model_description}

\begin{figure}[!t]
\includegraphics[width=8.5cm]{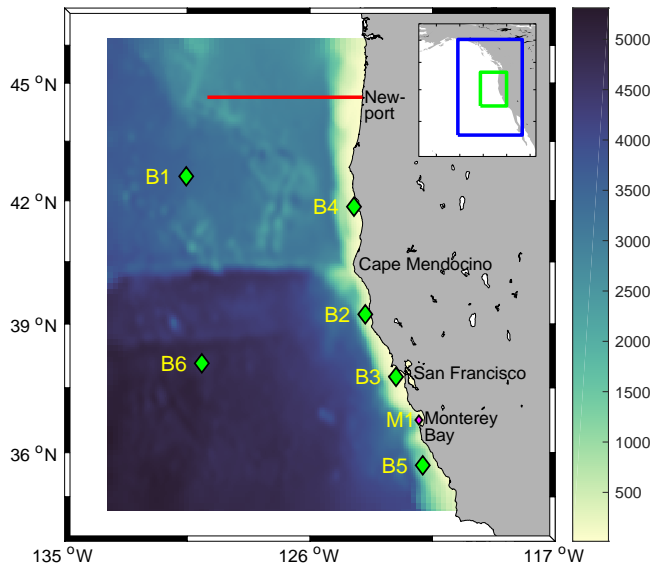}
\caption{The model area of the 3D modelling system applied to the CalUS. The small panel 
top right depicts the extent of the parent grid (blue box) and the one-way nested child 
grid (green box), whose bathymetry is depicted within the larger panel. The symbols B1 - 
B6 denote the NDBC buoys (used for temperature validation). The red line denotes the 
position of the Newport transect. Finally, the magenta symbol M1 denotes the position of the M1 
buoy of Monterey Bay Aquarium Research Institute (MBARI).}
\label{fig:fig_model_area}
\end{figure}

For studying the CalUS, we decided to use the ROMS modelling system 
\citep[\textbf{R}egional \textbf{O}cean \textbf{M}odelling \textbf{S}ystem, see 
e.g.][]{Haidvogel:2000, Wilkin:2005}, as it has been 
applied to this study region many times before. \\ Here, within this modelling system a 
3D coastal ocean circulation model is coupled to a lower trophic level nitrogen-based 
ecosystem model. In the following we will describe the two modules separately.


\subsection{Description of the hydrodynamic module}
\label{sec:note_hydrodynamic_module_description}

The hydrodynamic component is built using an online one-way nested system (see model 
domains in \Fig{fig:fig_model_area} ) with the coarser parent grid having a horizontal resolution 
of about 15 km and the nested child grid having a resolution of about 5 km, 
which corresponds to a three-times nesting refinement. In the vertical, 30 
layers of terrain-following S-coordinates are used with a strong refinement 
near the sea surface to resolve the upper 500 m of the water column and a smaller one near the sea floor (see 
the corresponding parameters in \Tab{tab:3d_model_hydro_parameter}). To capture subgrid-scale 
vertical turbulence, a k-kl variant of the Generic Length Scale (GLS) turbulence scheme 
\citep{Umlauf_Burchard:2003, WARNER_2005} is chosen for estimating vertical mixing coefficients for the 
hydrodynamic and the biological module, 
whereas for the horizontal mixing of momentum and active/passive tracers harmonic diffusion with constant 
turbulent diffusivities is used (see \Tab{tab:3d_model_hydro_parameter}). 
To deal with the turbulent vertical flux of horizontal momentum within the bottom boundary layer a quadratic 
bottom friction is selected.

In order to reduce the model spinup-time, initial conditions of free-surface elevation, horizontal water 
velocities, temperature and salinity for January 2012 are taken from the 1/12 degree global HYCOM + NCODA 
reanalysis data \citep[HYbrid Coordinate Ocean Model, see e.g.][]{Bleck:2002, Chassignet_al_2006}. This also 
means that there is a sufficient amount of turbulence already present at the beginning of the model 
simulation, which does not have to be built over a spinup process of several years as e.g. described by 
\citet{Marchesiello_2003}.

Hydrodynamic open-boundary conditions for the larger parent grid are obtained from two different data 
sources. In order to include tidal effects into the modelling system, tidal-harmonic constants are 
provided using the OSU Tidal Data Prediction Software\footnote{http://volkov.oce.orst.edu/tides/otps.html} 
together with the OTIS Regional Tidal Solutions dataset 'PO2009' for the Pacific Ocean 
\citep{Egbert_Erofeeva_OSU_2002}. In order to obtain an estimate of the true free-surface along the open 
boundary (tidal + mean background), free-surface elevation model data from the HYCOM + NCODA reanalysis  
are added to the obtained tidal signal. The vertical-mean horizontal velocities are obtained via the 
'reduced' boundary condition type from the prescribed ('clamped') free-surface elevation. However, as it is 
well known that the ROMS modelling system might face some problems at open boundaries when using these kind 
of 'reduced' boundary conditions, a new sponge-layer type is implemented into the ROMS model code to 
stabilize the model along the open boundaries. The working mechanism of this new sponge layer is explained in 
more detail within \ref{sec:sponge_layer}. The three-dimensional horizontal velocities are 'nudged' 
to velocity data obtained from the 1/12 degree global HYCOM + NCODA reanalysis data. With using these HYCOM + 
NCODA reanalysis data sets, the parent model domain is included into the large-scale hydrodynamic circulation 
of the Pacific Ocean.

\begin{table}[!b]
\caption{Model parameters used for the hydrodynamic module of the  3D coupled modelling system}
\begin{tabular}{ll}
\hline \hline
General parameters: & \\
& \\
Number of vertical layers & 30 \\
S-coordinate transformation equation & 2\\
S-coordinate stretching function & 4\\
S-coordinate surface control parameter & 7.0\\
S-coordinate bottom  control parameter & 0.5\\
Quadratic bottom drag coefficient. & 0.0025\\
3D velocity nudging time scale & 1.0 d\\
Tracer nudging time scale     & 1.0 d \\
Factor outflow/inflow nudging & 10 \\
& \\
\hline
Parent grid: &  \\ 
& \\
External/barotropic time step & 1.5 s\\
Internal/baroclinic time step & 30  s \\
Horizontal resolution & $\approx$ 15 km  \\
Horiz. turbulent viscosity & 300 m$^2$ s$^{-1}$ \\
Horiz. turbulent tracer diffusivity & 300 m$^2$ s$^{-1}$\\
& \\
\hline
Child grid: &  \\ 
& \\
External/barotropic time step & 0.5 s\\
Internal/baroclinic time step & 10  s \\
Horizontal resolution & $\approx$ 5 km  \\ 
Horiz. turbulent viscosity & 100 m$^2$ s$^{-1}$\\
Horiz. turbulent tracer diffusivity & 100 m$^2$ s$^{-1}$\\
\hline 
\end{tabular} 
\label{tab:3d_model_hydro_parameter}
\end{table}

\begin{figure}[!t]
\begin{center}
\includegraphics[width=8cm]{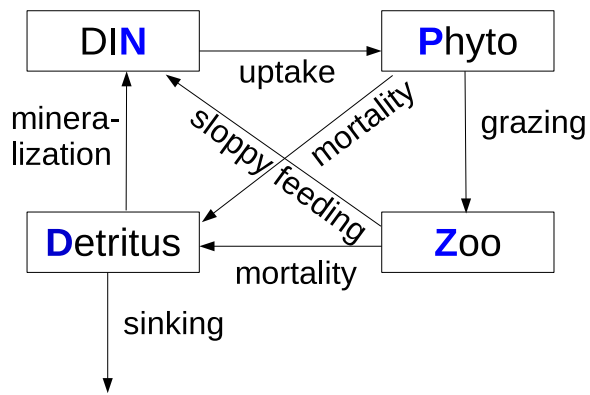} 
\end{center}
\caption{Schematic of the considered biological model after \citet{Powell_2006}:
dissolved inorganic nitrogen (N), particulate nitrogen (detritus:\,D), phototrophic 
phytoplankton (P), and herbivorous zooplankton (Z).}
\label{fig:NPZD_model_drawing}
\end{figure}

The atmospheric forcing (wind speed, air temperature, precipitation, sea-level pressure, total cloud 
cover and air humidity) is based on 3-hourly \citet{ERA5_2017} reanalysis  data with a spatial resolution of 
about 0.28 degree. 
Time-varying river runoff data for the year 2012 from USGS, USA\footnote{U.S. Geological 
Survey, 2016, National Water Information 
System data available on the World Wide Web (USGS Water Data for the Nation), accessed October 2018, at URL 
http://waterdata.usgs.gov/nwis/}, is also included for the following larger rivers: Stikine 
River (Alaska), Columbia River (Oregon / Washington), Rogue River (Oregon), 
Klamath River (Oregon / California), Eel River (California), Sacramento River 
(California), San Joaquin River (California). For each river, its discharge and 
contribution to salinity has been considered, whereas coastal temperatures and 
state variables of the biological module are not affected by river runoff.

\subsection{Description of the biological module}
\label{sec:3d_biological_module_description}

The \citet{Powell_2006} four-component NPZD model, which itself is mainly based on the 
studies by \citet{Spitz_2003} and \citet{Newberger_2003},  is used as a simple model with 
sufficient complexity to investigate biogeochemical conversion rates under the influence of turbulent 
transport processes. The model parameters are mainly taken as described within these articles (with one 
exception, see below). Therefore, we skip a very detailed validation of the biological 
model. Within this nitrogen-based trophic module,  total nitrogen is partitioned between 
dissolved inorganic nitrogen (N), particulate organic nitrogen (detritus:\,D), 
phototrophic phytoplankton (P), and herbivorous zooplankton (Z). 

The dynamics of each of these four components and their interactions are illustrated 
in \Fig{fig:NPZD_model_drawing} and can be described via a transport-reaction equation of 
the form: 
\begin{equation}
\p{C}{t} + \nabla \cdot (\vc{v} C) = \nabla_h \cdot (D_h \nabla_h C) +
\f{\partial}{\partial z}\left(D_v\p{C}{z}\right) + R
\label{eq:basic_transport_equation}
\end{equation}
where $\nabla$ denotes the nabla-operator, $\vc{v}$ 
the water velocity ($\vc{v} = (u,v,w)$), $D_h$ the horizontal eddy diffusion
coefficient, $D_v$ the vertical eddy diffusion coefficient, and finally, $R$ the net
conversion rate for each species.

\begin{figure}[!t]
\includegraphics[width=8.75cm]{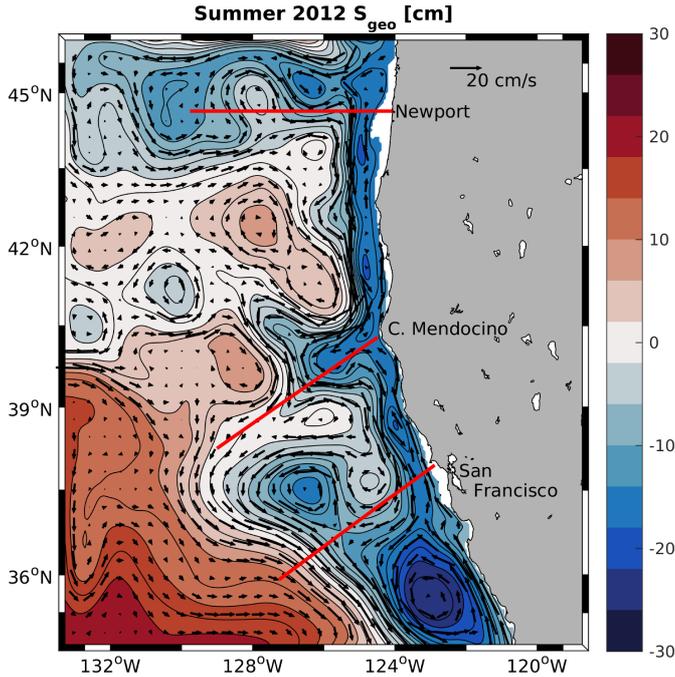}
\caption{The contour plot denotes the anomaly (the spatial mean has been subtracted) of 
the summer-mean modified geostrophic stream function $S_{geo}$ after \Eq{eq:app_S_geo} in 
100~m depth, whereas the summer mean Eulerian horizontal velocities at 100 m are denoted 
by the black arrows. The red lines mark the location of the vertical 
transects, for which normal velocities and temperatures are plotted within 
\Fig{fig:FIG_transect_data_ERA5_rivers_restart_1}.}
\label{fig:FIG_ERA5_1_summer_2012_S_geo}
\end{figure}

In detail, the net biological conversion rates for 
each species are \citep[for more details see][]{Powell_2006}:
\begin{eqnarray}
R_N & = & \delta D + \gamma_n G Z - UP \label{eq:bio_model_R_N}   \\ 
R_P & = & UP -G Z - \sigma_d P \\
R_Z & = & (1-\gamma_n) GZ - \zeta_d Z \\
\label{eq:3d_rate_term_D}
R_D & = & \sigma_d P + \zeta_d Z - \delta D ~~ + w_d \p{D}{z} 
\end{eqnarray}
The detritus rate term $R_D$ is augmented by the detritus sinking rate (the last term on the right in 
Eq.\,\ref{eq:3d_rate_term_D}), which is actually not a biological conversion term.
Furthermore, the following definitions are used:
\begin{eqnarray}
G & := & R_m (1 - e^{-\Lambda P})    \\ 
U & := & \f{V_m N}{k_U + N} \f{\alpha I}{\sqrt{V_m^2 + \alpha^2 I^2}}    \\ 
I & := & I_0 ~ par ~  exp\left(- k_z z - k_p \int_{0}^{z} P(z') dz' \right)  
                 \label{eq:bio_light_attenuation}
\end{eqnarray}
Light attenuation is modelled by \Eq{eq:bio_light_attenuation}, where $z$ 
denotes the (positive) vertical distance between the sea surface and the 
position within the water column, $I_0$ the variable sea-surface short wave 
radiation flux, and $par$ the fraction of light that is available for
photosynthesis \citep[see e.g.][]{Fennel_al_2006}.
The parameters of this NPZD-model are mainly identical to the values used in \citet{Powell_2006} and are 
listed in \Tab{tab:3d_model_bio_parameter}. However, the parameters of grazing
by zooplankton upon phytoplankton have been changed according to \citet{FIECHTER_2009}, as 
this leads to a larger mortality of phytoplankton due to grazing, which is more realistic 
for the CalUS (personal communication with Jerome Fiechter).

\begin{figure}[!t]
\includegraphics[width=8.5cm]{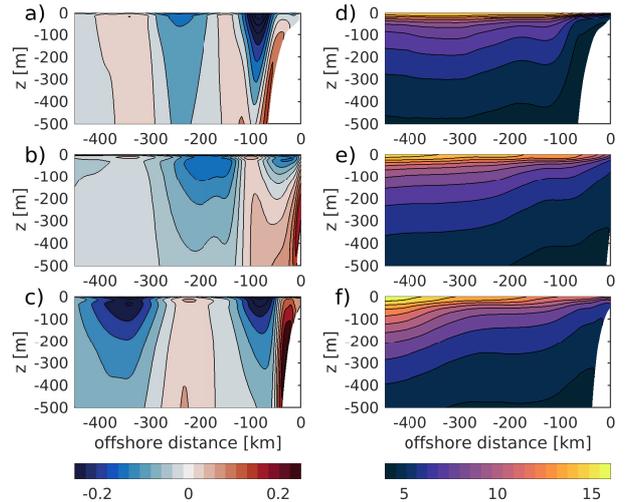}
\caption{Horizontal normal current velocity ([m s$^{-1}$]) (panels a) - c)) and potential 
temperature ([$ ^{o}$C]) (panels d) - f)) on the
vertical sections marked by the red lines in \Fig{fig:FIG_ERA5_1_summer_2012_S_geo}. Panel 
a) and d) correspond to the Newport transect, panel b) and e) to CalCOFI line 46.7 starting at Cape 
Mendocino, and panel c) and f) belong to CalCOFI line 60 starting at San Francisco. 
Normal velocities in northward direction are denoted by positive (red) values.}
\label{fig:FIG_transect_data_ERA5_rivers_restart_1}
\end{figure}

\begin{table}[!b]
\begin{small}
\caption{Model parameters used for the biological module of the coupled 3D modelling system \citep[mainly 
identical with][]{Powell_2006, Spitz_2003}.}
\begin{tabular}{lcl}
\hline 
Parameter Name &  Symbol &  Value \\
\hline
& &  \\
Light extinction coeff.       & $k_z$      & 0.067        m$^{-1}$\\
Self-shading coeff.           & $k_p$      & 0.0095       (m\,$\mu$M-N)$^{-1}$\\
Initial slope of P-I curve    & $\alpha$   & 0.025        m$^{2}$W$^{-1}$\\
PAR-fraction                  & $par$      & 0.43          \\
Phyto. max. uptake rate       & $V_m$      & 1.5          d$^{-1}$\\
Uptake half saturation        & $k_U$      & 1.0          $\mu$M-N \\
Phyto. senescence             & $\sigma_d$ & 0.1          d$^{-1}$\\
Zoop. grazing rate            & $R_m$      & 0.65         d$^{-1}$\\
Ivlev constant                & $\Lambda$  & 0.84         $\mu$M-N$^{-1}$\\
Excretion efficiency          & $\gamma_n$ & 0.3          \\
Zoop. mortality               & $\zeta_d$  & 0.145        d$^{-1}$\\
Remineralization              & $\delta$   & 1.03         d$^{-1}$\\
Detrital sinking rate         & $w_d$      & 8.0          m d$^{-1}$\\
\hline 
\end{tabular} 
\label{tab:3d_model_bio_parameter}
\end{small}
\end{table}

At the open boundaries of the model domain, a nudging method is used to force the 
biological variables to prescribed values. Apart from the DIN pool, all other 
biological variables are forced to zero for all times and all depth levels. 
However, dissolved inorganic nitrogen (DIN) is set according to the depth-dependent 
annual-mean \nitrate concentration obtained from the Levitus data set\footnote{Downloaded 
in March 2018 from http://iridl.ldeo.columbia.edu/\\SOURCES/.LEVITUS/index.html} 
\citep{Levitus_1982}. The initial conditions are set to a value of 1.0~mmol-N 
m$^{-3}$ for all variables apart from the nitrogen pool, which is set to 17.0~mmol-N 
m$^{-3}$. So, although DIN also includes other N-species like nitrite or amonium, the boundary and initial 
conditions were mainly set to nitrate based numbers.


\section{'Validation' of the modelling system}
\label{sec:modelling_system_validation}

After this technical description of the two coupled modules, we want to demonstrate that the modelling system 
at hand provides a sufficient first representation of the CalUS without reproducing all hydrodynamic and 
biogeochemical features of the CalUS.


\subsection{Validation of the hydrodynamic module}
\label{sec:3d_model_hydro_validation}

Within this part, the hydrodynamic module is to be validated by means of 
temperature data, data of free-surface elevation, and current velocity.

\begin{figure}[!t]
\includegraphics[width=8.5cm]{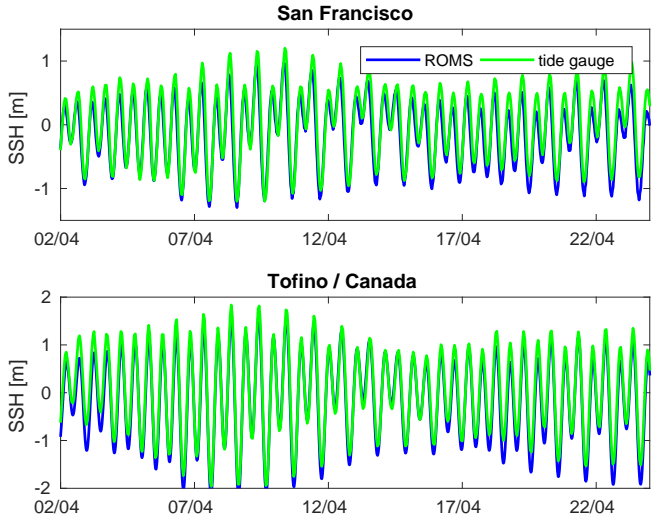}
\caption{Comparison of modelled sea surface height (SSH) with tide gauge data at Fort Point, 
San Francisco, and Tofino (Vancouver Island, Canada) over a time-period of two weeks in 
April 2012.}
\label{fig:pegel_validation}
\end{figure}

\begin{figure*}[!t]
\includegraphics[width=18.5cm]{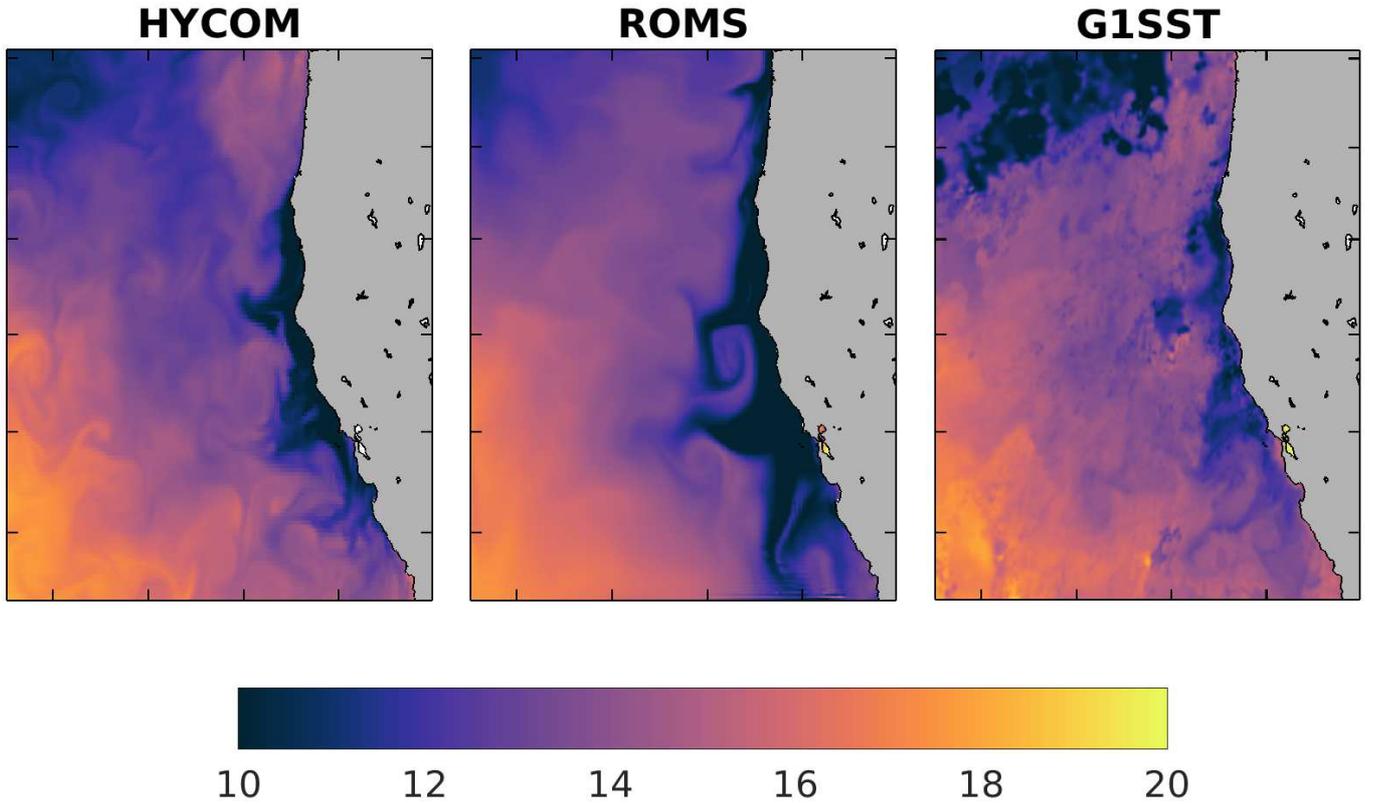}
\caption{SST at 16 June 2012 obtained from the 1/12 degree global HYCOM + NCODA 
reanalysis data (left), the hydrodynamic model component of the
ROMS modelling system presented in this study (middle), and finally from L4-G1SST satellite data (left).}
\label{fig:temp_validation_horizontal_with_sat}
\end{figure*}

\paragraph{Current Velocities}

First, we want to demonstrate that the model captures the main features of the horizontal circulation within 
the CCS as described above. As an example, \Fig{fig:FIG_ERA5_1_summer_2012_S_geo} depicts the summer 
mean\footnote{The mean is calculated as the mean of two-days averages over 90 days during the months 
June, July, and August 2012.} horizontal Eulerian velocity in 100 m depth. In addition, this figure also 
depicts the contour lines of the anomaly (the spatial mean has been subtracted) of the 2012 summer-mean 
modified dynamic height: 
\begin{equation}
S_{geo}(x,y,z) := \zeta(x,y) + \f{1}{\rho_0} \int_{z}^{\zeta} \rho_{ano}(x,y,z') dz'
\label{eq:app_S_geo}
\end{equation}
Here, $\zeta$ denotes summer-mean sea surface height, $\rho_0$~=~1000 
kg/m$^3$, $\rho_{ano}$ the density anomaly to 1000 kg/m$^3$ obtained from the summer mean distribution of 
potential temperature and salinity. As $g S_{geo} /f$ (with $g$ the gravitational acceleration, $f$ the 
Coriolis parameter) is a geostrophic streamfunction, $S_{geo}$ can be considered as some kind of scaled 
geostrophic streamfunction, whose contour lines should also be parallel to the geostrophic velocity.

As is evident from \Fig{fig:FIG_ERA5_1_summer_2012_S_geo}, the mean horizontal current velocities are 
more or less parallel to the streamlines, which is due to the validity of the geostrophic approximation. This 
picture shows the drop of $S_{geo}$ towards the coast; and the resulting coastal jet mentioned above is 
clearly visible (especially between Newport and Cape Mendocino). Similar to observations, this jet shows 
strong meandering, and separates from the coast south of Cape Mendocino as indicated in Fig. 3. of  
\citet{GANGOPADHYAY_al:2011}. While separating from the coast near Cape Mendocino, the coastal jet seems 
to split into two branches: one moving offshore, one flowing further near the coast. Also similar with 
this figure, within the south-west corner of \Fig{fig:FIG_ERA5_1_summer_2012_S_geo}, meanders and jets of the 
broader and slower CC seem to be visible.
In addition, there are also some strong closed contours of the summer mean $S_{geo}$, that might indicate 
standing eddies within that region. Some of them might corresponds to mentioned features within the 
literature, as the one south of San Francisco near $36$ N, that might correspond to the San Francisco eddy 
mentioned in \citet{HICKEY:1979}; likewise the smaller one near Point Arena (near 39 N) mentioned in 
\citet{Hayward_Mantyla:1990}. 

The mentioned hydrodynamic features are also visible on vertical sections of the horizontal current 
velocity depicted in \Fig{fig:FIG_transect_data_ERA5_rivers_restart_1}, and the positions of these 
transects of 450 km length are marked by the red lines in \Fig{fig:FIG_ERA5_1_summer_2012_S_geo}. The 
northern most transect is chosen according to the transect presented in  \citet{Powell_2006}. The two 
southern lines are chosen after the  California Cooperative Fisheries Investigations (CalCOFI) sampling grid. 
The line starting at Cape Mendocino corresponds to CalCOFI transect with line coordinate\footnote{The 
transformations between the CalCOFI sampling coordinates and geographic coordinates in latitude and longitude 
are performed after \citet{Weber_Moore:2013} using the Matlab software package by Robert Thombley and 
Augusto Valencia downloaded from 
http://calcofi.org/field-work/station-positions/calcofi-line-sta-algorithm.html.} 46.7, whereas the southern 
most line starting at San Francisco corresponds to line coordinate 60. Within a 100 km band close to the 
coast, the coastal jet is clearly visible in \Fig{fig:FIG_transect_data_ERA5_rivers_restart_1} a) and c). 
Near Cape Mendocino,  due to a meander, the jet has moved more offshore and is visible 200 km from the coast. 
From this figure, the vertical extent of the main core of the coastal jet is clearly visible to be within the 
upper 200~m. Furthermore, on all three transects, the California Undercurrent (CU) is visible as the red band 
near the continental slope with its core to be between 400 - 200~m. However, the CU reaches upwards to 100 m, 
such that its northward flow velocity is also visible by the northward pointing velocity vectors close to the 
coast in \Fig{fig:FIG_ERA5_1_summer_2012_S_geo}.

\begin{table}[!b]
\begin{centering}
\caption{Results of the tidal analysis for year 2012 obtained via the T-tide software package 
\citep{Pawlowicz:2002} at San Francisco and Tofino (Vancouver Island, Canada) tide-gauge stations. The 
phase difference is obtained by subtracting the phase value of the modelled data from the phase 
value of the measured data. So, a negative phase difference denotes the model lagging behind the data.}
\begin{tabular}{lcccc}
\hline
  & \multicolumn{2}{c}{San Francisco} & \multicolumn{2}{c}{Tofino}   \\ 
  & data  & model & data & model  \\ 
\hline 
\hline
$M_2$ amp.  [m]  & 0.57 &0.52  & 0.97 & 0.96  \\ 
$M_2$ $\Delta \phi$ [$^\circ$] & \multicolumn{2}{c}{15.05} & \multicolumn{2}{c}{-0.50} \\ 
$M_2$ $\Delta \phi$ [min] & \multicolumn{2}{c}{31.15} &  \multicolumn{2}{c}{-1.04} \\ 
\hline
$S_2$ amp.  [m] & 0.13 &0.12  & 0.28 &  0.29 \\ 
$S_2$ $\Delta \phi$  [$^\circ$] & \multicolumn{2}{c}{10.0} & \multicolumn{2}{c}{-3.72}\\ 
$S_2$ $\Delta \phi$  [min] & \multicolumn{2}{c}{19.99} &  \multicolumn{2}{c}{-7.43} \\ 
\hline
$K_1$ amp.  [m] & 0.37 &0.41  & 0.39 & 0.49  \\ 
$K_1$ $\Delta \phi$  [$^\circ$] & \multicolumn{2}{c}{8.94} & \multicolumn{2}{c}{6.57}\\ 
$K_1$ $\Delta \phi$  [min] & \multicolumn{2}{c}{35.66} &  \multicolumn{2}{c}{26.20} \\ 
\hline
$O_1$ amp.  [m] & 0.23 &0.26  & 0.24 &  0.31 \\ 
$O_1$ $\Delta \phi$   [$^\circ$] & \multicolumn{2}{c}{8.07} & \multicolumn{2}{c}{7.24}\\ 
$O_1$ $\Delta \phi$  [min] & \multicolumn{2}{c}{34.72}  &  \multicolumn{2}{c}{31.15} \\ 
\hline 
\end{tabular} 
\label{tab:tidal_analysis_parameter}
\end{centering}
\end{table}

\paragraph{Sea Surface Height (SSH)}

As the tides quite significantly contribute to the short-term variability, we want to pay 
some attention to their representation within the model. As the model is forced along the 
open boundaries by tidal elevations as described above (here, model errors are only due 
to the errors within the forcing data), it might be questionable if the (tidal) 
fluctuations of SSH in the central part of the model domain are of the 
right order. Therefore, tide gauge data of two stations located in the central area of 
the model domain have been considered for model validation: from Tofino (Vancouver 
Island, Canada) within the coarser grid, and from San Francisco (Fort Point) within the 
finer child grid. As a first inspection, SSH time series for a 
two-weeks period in April 2012 are depicted in \Fig{fig:pegel_validation}. Although not matching perfectly, 
it is evident from this figure that the tides are represented 
appropriately with respect to amplitude and phase at these two tide gauge stations. This 
is further confirmed by using a Taylor diagram analysis (see 
\Fig{fig:taylordiagram_buoy_gauge_validation}) for the total year 2012, which shows the 
good agreement in terms of standard deviation and correlation. In addition, the harmonic 
analysis (using the T-tide software package of \citet{Pawlowicz:2002}) presented in 
\Tab{tab:tidal_analysis_parameter} shows a satisfactory agreement of the tidal 
amplitudes of the harmonics M2, S2, O1, and K1 for both tide gauge stations. However, 
there is some phase lag in the order of some minutes of the M2 and S2 tide in San 
Francisco, which might be due to small errors in tidal wave propagation. In the end, this 
validation process demonstrates that the tidal dynamics in the CalUS is represented to 
the right order of magnitude within the hydrodynamic module at hand.


\paragraph{Water Temperatures}

\begin{figure}[!t]
\includegraphics[width=8.5cm]{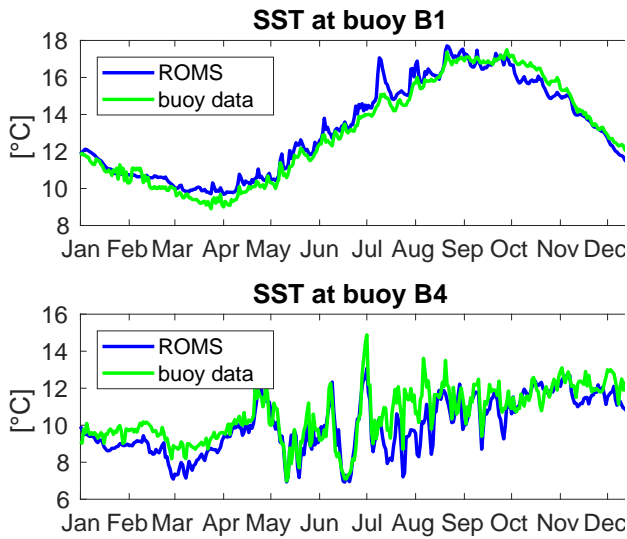}
\caption{Comparison of modelled (blue line) and measured (green line) sea surface 
temperature (SST) at buoys B1 and B4  
for the year 2012 (for the location of these buoys, please see \Fig{fig:fig_model_area}).}
\label{fig:temp_buoy_validation}
\end{figure}

As signatures of upwelling processes are usually visible as specific features of sea 
surface temperature, we compare time series of modelled and measured SST at different sites within the CalUS, 
for which buoy data have been available from the NDBC\footnote{National Data Buoy Centre, www.ndbc.noaa.gov}. 
The locations of these buoys are shown in \Fig{fig:fig_model_area} and are denoted by symbols B1 - B6, 
which correspond to the following buoy numbers: B1 - 46002; B2 - 46014 ; B3 - 46026 ; B4 - 46027; B5 - 46028; 
B6 - 46059.

\Fig{fig:temp_buoy_validation} depicts the 2012 time series of SST at two buoy sites: one 
close to the coast 
(B4) and one more offshore (B1). This figure demonstrates that the model captures the 
long-term 
evolution of the SST quite well. It even reproduces the timing and strength of some 
upwelling events in May 
and June 2012, which can be seen from the lower panel in \Fig{fig:temp_buoy_validation}. 
However, is is also 
visible from the time series of the near-shore site that the model might overestimate the 
strength of the upwelling process a bit, which could explain the underestimation of the 
near-shore SST. 
In addition, the SST comparisons for the other buoy sites are depicted within the Taylor 
diagram in 
\Fig{fig:taylordiagram_buoy_gauge_validation}, from which it is evident that the SST 
dynamics is captured by 
the model in a sufficient manner.

\begin{figure}[!t]
\includegraphics[width=8.5cm]{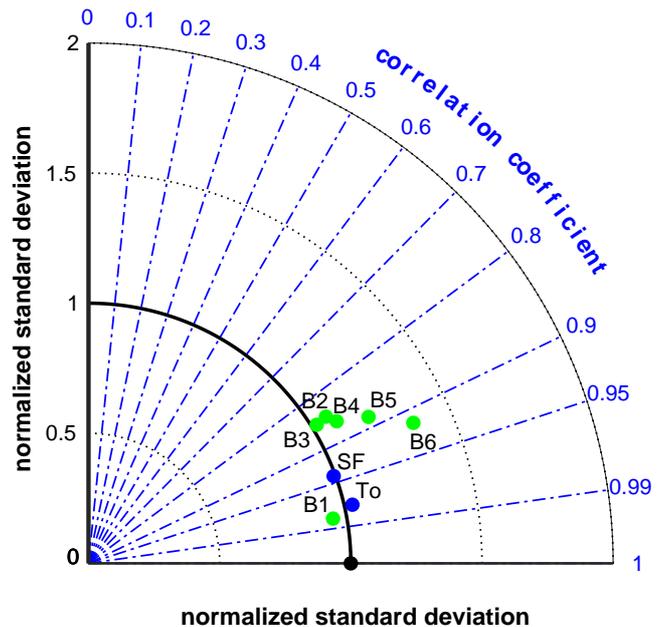}
\caption{Taylor diagram for validation of modelled sea surface temperature at buoys B1 - 
B6 
(for the location of these buoys, please see \Fig{fig:fig_model_area}) and SSH at tide 
gauge stations located in San Francisco (SF) and Tofino (To) (Vancouver Island, Canada).}
\label{fig:taylordiagram_buoy_gauge_validation}
\end{figure}

This impression of a slight overestimation of the upwelling process can also been reasoned from 
\Fig{fig:temp_validation_horizontal_with_sat}, which shows a spatial plot of modelled SST for 16 June 2012 
together with SST fields obtained from the HYCOM model and satellite data. Although the position and extent 
of the upwelling region are quite similar to those features within the HYCOM data and the L4-G1SST satellite 
data\footnote{Data were obtained via the web-portal worldview.earthdata.nasa.gov.}, our model seems to 
overestimate the upwelling process in that time period to some extent, which is evident from the colder SST 
values near the coast. However, looking at special features of SST at that day, the model clearly shows 
filaments as well as mushroom-like and eddy structures due to the upwelling process as mentioned above.

In addition, the near-shore upwelling of colder subsurface waters can also be seen on the 
vertical CalCOFI sections depicted in \Fig{fig:FIG_transect_data_ERA5_rivers_restart_1}~d)~-~f), which are 
comparable to vertical temperature sections shown in \citet{Marchesiello_2003} or 
\citet{Chenillat_al_2013}.


\subsection{'Validation' of the biological module}
\label{sec:3d_biological_module_validation}

\begin{figure}[!t]
\includegraphics[width=8.5cm]{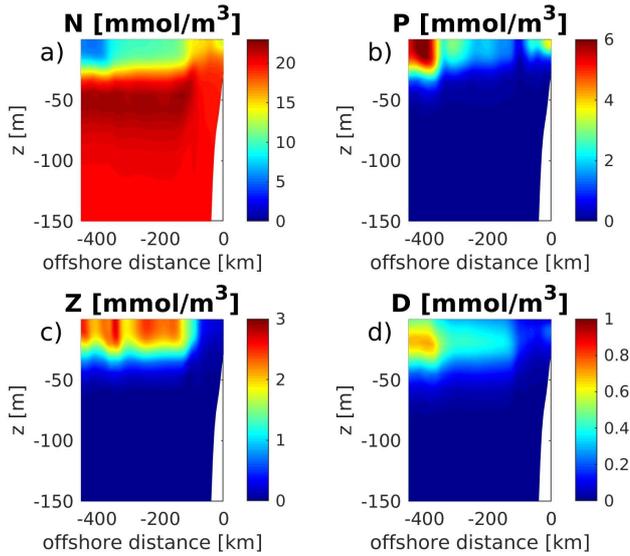}
\caption{Monthly mean values for May 2012 of the four species of the biological 
model on the vertical transect near Newport.  The location of this transect is depicted 
within \Fig{fig:fig_model_area}.}
\label{fig:FIG_transect_bio_may}
\end{figure}

To illustrate the performance of the biological module, monthly mean values of dissolved 
inorganic nitrogen (DIN), phytoplankton and zooplankton are plotted along the Newport 
transect \citep[similar to ][red line in \Fig{fig:fig_model_area}]{Powell_2006} for May 2012 in 
\Fig{fig:FIG_transect_bio_may}. The upwelling of DIN-rich water mass is visible at the 
coast in panel \Fig{fig:FIG_transect_bio_may}a, as well as the DIN consumption within 
the euphotic zone by growing phytoplankton.  The growing of zooplankton is visible in 
panel \Fig{fig:FIG_transect_bio_may}c, and  the remineralization of detritus to DIN 
is evident from the increase of nutrients below the euphotic zone within the 
depth-interval between 50 - 100 m in \Fig{fig:FIG_transect_bio_may}a.

\begin{figure}[!t]
\includegraphics[width=9cm]{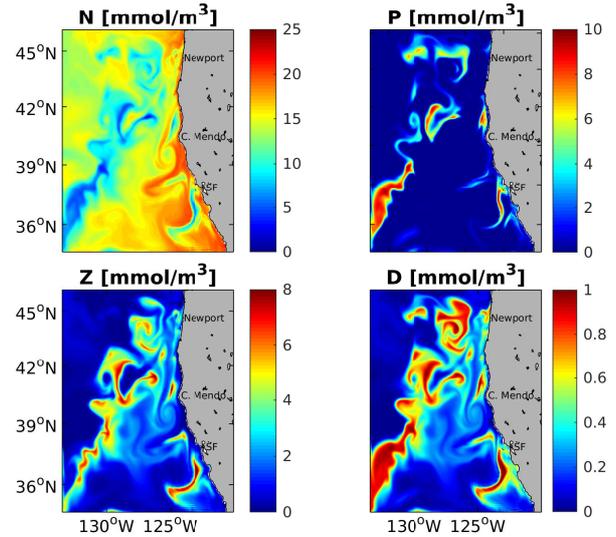}
\caption{Exemplary horizontal distributions of the biological module at 16 June 2012 in a 
water depth of 15~m below sea surface.}
\label{fig:FIG_bio_fields_horizontal}
\end{figure}

The interaction and the temporal succession of the N-species is also visible within a temporal snapshot of 
near-surface distributions of the N-species depicted in \Fig{fig:FIG_bio_fields_horizontal}. 
This figure shows the horizontal mesoscale structure of the biological variables, which is 
similar to the structures depicted in \citet{Powell_2006}, and which also shows an increase in nutrients 
towards the coast due to the prevailing upwelling dynamics. As an example, within the south-western region of 
the four panels of this figure, a stretched structure is visible as a depletion of N, which is also present 
as a build-up within the other panels for P, Z and D. Thus the temporal succession of nitrogen though the 
four pools is visible.

\begin{figure}[!t]
\includegraphics[width=9cm]{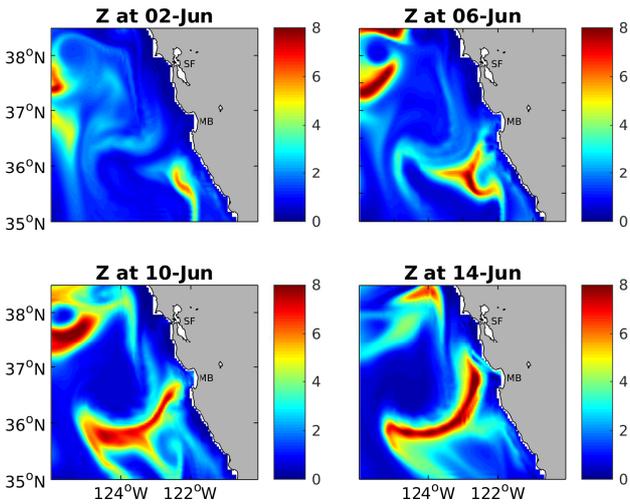}
\caption{Exemplary horizontal evolution of a zooplankton [mmol/m$^3$] patch in a depth of 
15~m below sea surface in the first half June 2012.}
\label{fig:FIG_Zoo_patch_horizontal}
\end{figure}

What is also visible within these horizontal distributions, is the patchy structure of the zooplankton 
species that is described e.g. in \citet{Messie_al_2017}. \citet{Messie_al_2017} and 
\citet{Fiechter_2020} investigate the formation and occurrence of zooplankton hotspots 
within the CalUS. They attribute their 
formation and distribution to regions of coastal nutrient upwelling as well as converging and 
diverging surface currents. The simple NPZD model at hand shows, to some extent, a similar behaviour as 
depicted in  \Fig{fig:FIG_Zoo_patch_horizontal}. Within that figure, the temporal evolution a  zooplankton 
patch near Monterey Bay (denoted as MB within that figure) is shown over a time period of nearly two weeks in 
June 2012. The simultaneous generation of zooplankton from phytoplankton within an upwelling centre as well 
as the horizontal transport of zooplankton within a narrow stripe to offshore locations is visible from this 
figure.

In order to investigate and validate the temporal dynamics of this simple NPZD model, modelled 
near-surface DIN concentrations are compared to measured nitrate values at a long-term time series station 
in Monterey Bay (M1 buoy, see \Fig{fig:fig_model_area} for the position of this buoy) 
\citep{Sakamoto_al_2017,Chavez_al_2017}. Concerning the origin of upwelled water masses,  Monterey Bay is 
mainly influenced by the upwelling center off A\~{n}o Nuevo (to the north of the bay), from which cold and 
nutrient-rich water enters Monterey Bay \citep[see e.g.][]{Chavez_al_2017}, although upwelling and/or mixing 
occurs along the entire region from A\~{n}o Nuevo to Point Sur. 
Due to its wind-protection capabilities, its slower circulation and a warm and stable mixed layer, Monterey 
Bay is a classical upwelling shadow environment that foster dense phytoplankton blooms  
\citep{Chavez_al_2017}.

\begin{figure}[!t]
\includegraphics[width=9cm]{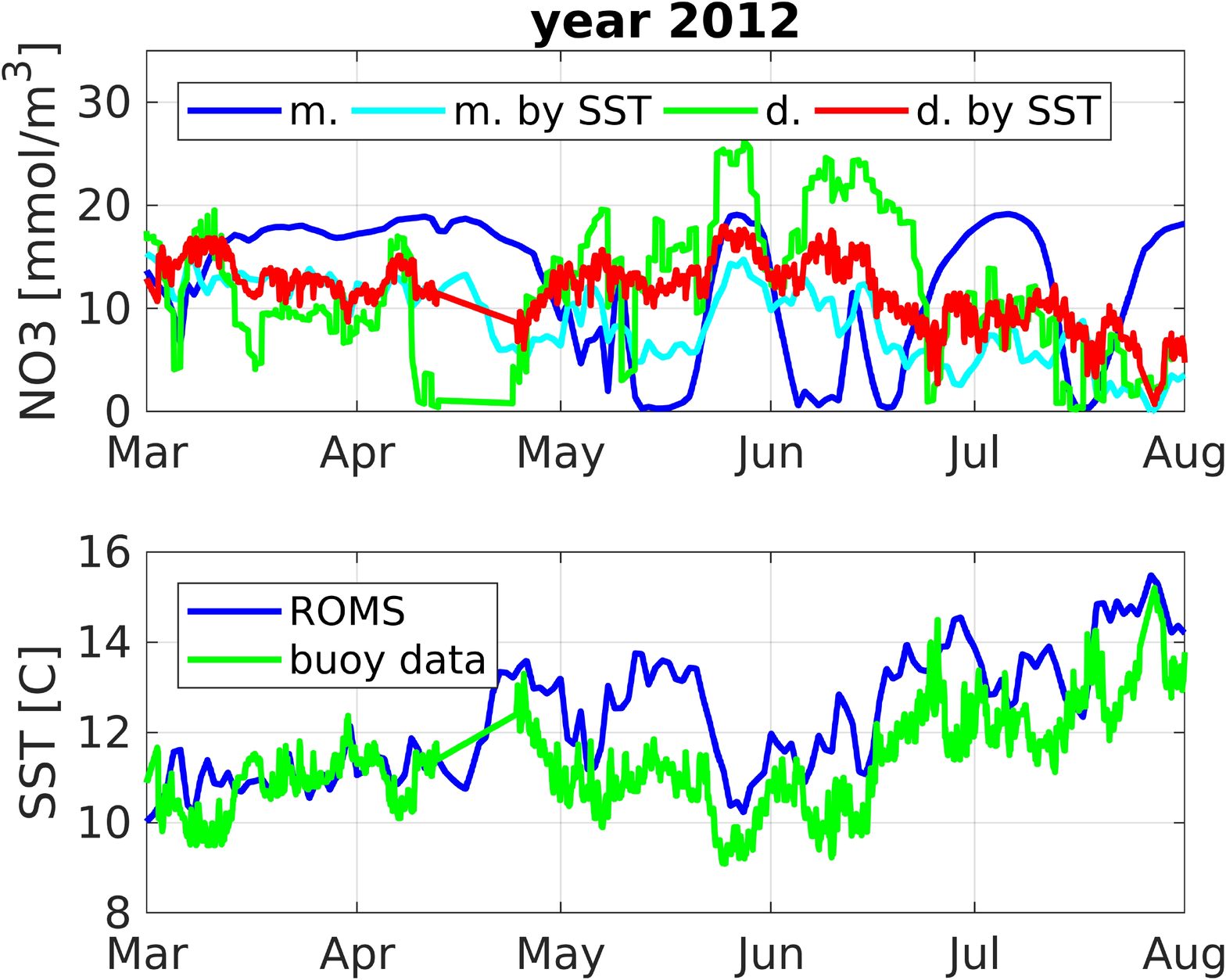}
\caption{Time series of modelled and measured nitrate concentrations (upper panel) as well as sea surface 
temperature (SST, lower panel) at MBARI buoy M1 in 2012. (See \Fig{fig:fig_model_area} 
for the position of this buoy in Monterey Bay.) Within the upper panel, the blue line denotes the modelled 
(m.) 'nitrate' concentration, whereas the measured is denoted by the green line (d.). Using a simple 
linear regression model based on measured SST, the other two lines denote the estimated nitrate 
concentration using \Eq{eq:no3_sst_regression_model} (m. by SST - using modelled SST; d. by SST - using 
measured SST).}
\label{fig:FIG_time_series_NO3}
\end{figure}

As mentioned above, DIN includes more N-species apart 
from nitrate, such that a direct comparison of nitrate and DIN is problematic, However, the DIN is 
initialized and set at open boundaries to the order of long-term nitrate concentrations. 
Therefore, the DIN within this manuscript is of similar magnitude as nitrate.
According to \citet{Chavez_al_2017} the nitrate concentration at M1 buoy at the entrance of Monterey Bay is 
strongly influenced by coastal upwelling of deep nutrient rich waters. Riverine nitrate input is negligible 
at that site in general and only of some importance during winter months \citep{Sakamoto_al_2017}. The 
long-term (1988 - 2016) climatological nitrate concentration shows a peak due to coastal upwelling during 
March and July \citep{Chavez_al_2017}, which is to some extent visible for the year 2012 in 
\Fig{fig:FIG_time_series_NO3} (see green line in upper panel). It is evident from this figure that the 
modelled nitrate concentration (blue line in upper panel) is of similar magnitude. However, it is also evident 
that the two time series (modelled and measured) do not match. As an example, the modelled time series depicts 
some peak at the beginning of August 2012, whereas the measured nitrate data show a falling trend with low 
absolute values. 

In the lower panel of \Fig{fig:FIG_time_series_NO3}, the modelled (blue line) and measured (green line)
SST is depicted, which shows that the overall trend and the right order of magnitude is captured by the 
model. However, some the short-term fluctuations are not always captured by the model, as can e.g. be seen in 
mid-May as well as at the beginning of July, when some cooler (an likely upwelling) events are missed by the 
model. 

In order get some idea, what causes these differences, we also compared the SST at this 
site, which seems to 
be a good indicator of upwelling waters. According to \citet{Sakamoto_al_2017}, as a first approximation, the 
nitrate concentration (mmol/m$^3$) can be estimated from sea surface temperature (SST, $^o$C) by a simple 
linear regression model (correlation coefficient $r^2$~=~0.59) of the form:
\begin{equation}
\text{NO}_{3}^- =  -2.82 \times \text{SST} +  43.59
\label{eq:no3_sst_regression_model}
\end{equation} 
Thus, the top panel also shows the SST-based estimated nitrate concentrations using the modelled (cyan line) 
and measured (red line) sea surface temperature at buoy M1.

It is evident that the model seems to capture an upwelling event with the largest measured nitrate 
concentration at the end of May, which is visible in all the four nitrate curves. However, the model shows 
high measured nitrate concentration at the beginning of July and end of July, which only partially seem to be 
related to modelled upwelling events. Whereas the event at the beginning of July shows some modelled 
upwelling signature (the cyan curve shows a small peak), the high nitrate concentration at the end of July / 
beginning of August has no upwelling counterpart.

The reason for the occasional overestimation of the modelled nitrate data might 
be due to the initialization of the nitrate concentration within the model domain, which also puts hight 
nutrient loads in the upper waters of the euphotic zone which is evident from \Fig{fig:FIG_transect_bio_may}, 
which shows nutrient concentration of about 15 mmol/m$^3$ in offshore waters. Therefore, in the model, in 
addition to upwelling of nutrients from below, horizontal currents could transport nutrients to buoy M1, 
which might explain the nitrate peaks outside the upwelling season. Alternatively, the temporal dynamics of 
the uncoupled (i.e. without any spatial transports) NPZD model could be visible on this figure.

In order to improve the model, the initialization of nutrients should be done in that way that the surface 
waters down to a depth of about 200~m should be free of nutrients. Only deeper waters should contain initial 
nutrient loads that can be brought to the surface by coastal upwelling. As an alternative, the model 
should have spun up over several years to deplete offshore waters from nutrients due to detrital sinking.

\section{Summary and conclusions}
\label{sec:conclusions}

A coupled modelling system for the California Upwelling System has been presented and validated. 
The hydrodynamic performance of the model is investigated in more detail by means of a tidal analysis 
against tide gauge data, a comparison of modelled sea surface temperature (SST) against buoy and satellite 
data, as well as vertical sections of along-shore currents and water temperature.
Although the upwelling dynamics along the coast might be a bit overestimated as demonstrated by means of 
SST, in the end, the validation process demonstrates that the hydrodynamic module used 
within this study is capable to reproduce the basic hydrodynamic and circulation features 
within the CalUS.

However, the usage of the simple NPZD module for modelling the real nutrient
and planktonic dynamics in that region is quite conceptual, 
as more sophisticated models are standard and available \citep[see 
e.g.][]{Fiechter_2020}. On the other hand, the simple NPZD model at hand is able to show some basic features 
of plankton dynamics with the right order of magnitude within that region. 

To conclude, the presented modelling system might be a valuable tool to investigate the physical and (to some 
basic extent) the biogeochemical dynamics within the CalUS. However, although the physical module already 
seems to be 'matured' to a sufficient degree, the biological has to be improved by means of more realistic 
initial and boundary conditions as well as a more sophisticated structure of the underlying 'food-web' (e.g.  
using more planktonic species and higher trophic levels).

\section{Acknowledgments}

We are grateful to Klemens Buhmann, Matthias Schr\"o-der and Stefan Harfst for technical support. 
We want to thank J\"org-Olaf Wolff and David M. Checkley Jr. for  help and advice concerning 
the manuscript. 
The numerical simulations were performed on the high performance computing cluster 
CARL, financed by the Ministry for Science and Culture (MWK) of Lower Saxony, Germany, and 
the German Research Foundation (DFG). 
VS was funded within the research project 'ENVICOPAS - Impact of Environmental Chances on 
Coastal Pathogen Systems (ENVICOPAS)' by German Research Foundation (DFG) under grant 
number 283700004. FH was funded within the research project 'Macroplastics Pollution in the 
Southern North Sea - Sources, Pathways and Abatement Strategies' by the Ministry for Science and Culture 
(MWK) of Lower Saxony, Germany. 

We finally want to mention some of our data sources:
The 1/12 deg global HYCOM + NCODA Ocean Reanalysis was funded by the U.S. Navy and the Modeling and 
Simulation Coordination Office. Computer time was made available by the DoD High Performance Computing 
Modernization Program. The output is publicly available at http://hycom.org.
Atmospheric forcing  data for estimating the upwelling index were obtained from NCEP Reanalysis provided by 
the NOAA/OAR/ESRL PSD, Boulder, Colorado, USA, \citep[see e.g.][]{Kalnay_NCEP:1996}.
The L4-G1SST satellite data set, a blended Global 1-km Sea Surface Temperature Data Set for Research and 
Applications was provided by Yi Chao, Benyang Tang, Zhijin Li, Peggy Li, Quoc Vu, Jet Propulsion Laboratory.
The atmospheric forcing for the hydrodynamic modelling system contains modified Copernicus Climate Change 
Service Information for the year 2012 provided by 'Copernicus Climate Change Service (C3S) (2017): ERA5: 
Fifth generation of ECMWF atmospheric reanalyses of the global climate. Copernicus Climate 
Change Service Climate Data Store (CDS), downloaded in August 2018 via \\ 
https://cds.climate.copernicus.eu/cdsapp.

Last but not least, we want to say 'thank you' to  Monterey Bay Aquarium Research Institute (MBARI) (by name 
Reiko Michisaki) for providing the ISUS nitrate time series data \citep{Chavez_al_1994} obtained at their 
surface buoy M1. MBARI provides data "as is", with no warranty, express or implied, of the quality or 
consistency. 
Data are provided without support and without obligation on the part of the Monterey Bay Aquarium Research 
Institute to assist in its use, correction, modification, or enhancement.

\appendix



\section{The proposed sponge layer type}
\label{sec:sponge_layer}

A new type of sponge layer is introduced into the ROMS source code to stabilize the 
vertical-mean velocity at open boundaries when so-called reduced boundary conditions are used. In this case, 
the free-surface elevation is prescribed at open-boundary points and the vertical-mean velocities are 
derived from a simplified momentum balance including pressure and Coriolis force. As this single prescription 
of the free-surface elevation and the derivation of the vertical-mean velocity might not be a consistent 
boundary condition, it is observed quite often that the model becomes unstable at open-boundary points. 
Therefore, a new type of sponge layer has been introduced into the ROMS source code, which has also been 
implemented into the unstructured-grid ocean model FVCOM \citep[see e.g.][]{Chen_al:2003, Chen_al:2007, 
Qi_FVCOM_SWAVE_2009} as explained in \citet{Kirchner_al:2020}, and which we want to 
explain here for completeness, as well.

\begin{figure}[t]
\includegraphics[width=8.5cm]{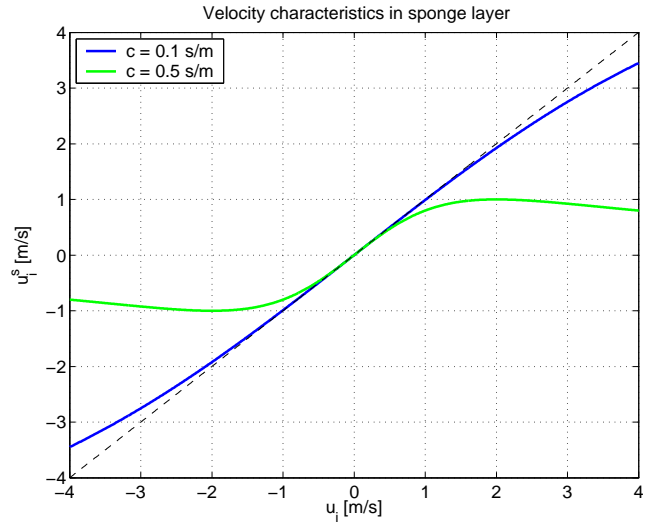}
\caption{Illustration of the sponge layer characteristics used at open-boundary points to stabilize the 
vertical-mean velocity when using the sponge layer in combination with 'reduced' boundary conditions. The 
green and blue line correspond to sponge-layer curves for $c\neq 0$. The black dashed line denotes the 
sponge-layer curve for $c=0$.}
\label{fig:sponge_layer_characteristics}
\end{figure}

To stabilize the vertical-mean horizontal velocity components $(u,v)$ at open-boundary grid points, each 
velocity component is modified according to the following equation:
\begin{equation}
u^s_i = \f{u_i}{ 1 + [c ~ u_i]^2}
\label{eq:sponge_layer_1}
\end{equation}
$u_i$ denotes the eastward or the northward component, $c$ denotes the sponge-layer parameter, with 
$c=0$ switching-off the sponge layer. And $u^s_i$ denotes the velocity value being further used at the open 
boundary. 

In \Fig{fig:sponge_layer_characteristics} two sponge-layer curves are depicted for two different 
sponge-parameter values. It is evident from this figure that the modification of the vertical-mean velocity 
along the open boundary is very small for small velocity values. In this case, the velocity is not 
affected by the sponge-layer. Depending on the value of the  $c$-parameter, the sponge layer is only 
effective for larger velocity values which e.g. might be encountered in case of instabilities.

This sponge-layer characteristic after \Eq{eq:sponge_layer_1} is symmetric around zero, which might be a 
disadvantage in case of valid currents at open boundaries which enter or leave the model domain. In 
this case, the mean velocity of the open-boundary currents might be decreased too much by the sponge-layer. 
In order to solve this issue, one could use a slightly modified version of \Eq{eq:sponge_layer_1}. Suppose 
the model should meet a background current velocity at the open boundary denoted by $u^b_i$, then the 
following equation dampens differences to this background velocity wanted:
\begin{equation}
u^s_i = \f{u_i}{ 1 + [c ~ (u_i - u^b_i)]^2}
\label{eq:sponge_layer_2}
\end{equation}
Again, within this study, the sponge-layer after \Eq{eq:sponge_layer_1} is used.

\bibliographystyle{elsarticle-harv}
\bibliography{CalUS_model_system_B.bib}

\end{document}